\documentclass[fleqn,usenatbib]{mnras}

\usepackage{newtxtext,newtxmath}

\usepackage[T1]{fontenc}
\usepackage{ae,aecompl}

\usepackage{graphicx,epic}	
\usepackage{amsmath}	
\usepackage{amssymb}	
\usepackage{hyperref}

\usepackage[normalem]{ulem}  
\usepackage{color,soul}

\def\psr{PSR~J2052+1219}
\title[The black widow PSR J2052+1219]{Optical detection of the black widow binary PSR J2052+1219}

\author[S. Zharikov,  A. Kirichenko,  D.  Zyuzin et al.]{
S. Zharikov\,$^{1}$\thanks{E-mail: zhar@astro.unam.mx},
A. Kirichenko$^{1,2}$,
D.  Zyuzin$^{2}$,
Yu.  Shibanov$^{2,3}$
and J. S. Deneva$^4$ 
\\
$^{1}$Instituto de Astronom\'ia, Universidad Nacional Aut\'onoma de M\'exico, Apdo. Postal 877, Ensenada, Baja California, M\'exico, 22800\\
$^{2}$Ioffe Institute, 26 Politekhnicheskaya st., St. Petersburg 194021, Russia\\
$^{3}$Peter the Great St.~Petersburg Polytechnic University, Politekhnicheskaya 29, St. Petersburg, 195251, Russia \\
$^{4}$George Mason University, resident at the Naval Research Laboratory, 4555 Overlook Ave. SW, Washington, DC 20375, USA
}

\date{Accepted  2019 August 30. Received 2019 August 12, in original form 2019 May 21}

\pubyear{2019}

\begin{document}
\maketitle

\begin{abstract}
We present optical  time-resolved multi-band photometry
of the black widow binary millisecond pulsar J2052+1219 using direct-imaging observations 
with the 2.1m telescope of Observatorio Astronomico Nacional San Pedro M\'artir, Mexico (OAN-SPM).
The observations revealed a variable optical source whose position and periodicity $P=2.752h$ 
coincide with the pulsar coordinates and the orbital period obtained from radio timing. This allowed us to identify it with the binary companion of the pulsar.
We reproduce light curves of the source  
modelling  the companion heating by the pulsar and accounting for the system  parameters obtained  from the radio data. As a result,
 we independently estimate the distance to the system of 3.94(16) kpc, which  agrees with the dispersion measure distance.
The companion star size is 0.12-0.15~R$_{\sun}$, close to filling its Roche lobe. It has a surface temperature 
difference of about 3000~K  between the side  
facing the pulsar and the back side. We summarise characteristics of all black widow systems studied in the optical and 
compare them with the PSR J2052+1219 parameters derived from our observations.

\end{abstract}




\section{Introduction}
\label{Intro}

Millisecond pulsars (MSPs) form a class of old neutron stars (NSs) that are characterised by short and stable rotational periods $P$ with typical $P < $~30 ms and 
$\dot{P}\la 10^{-19}$ s~s$^{-1}$. It is generally accepted that MSPs are ``recycled'' by accretion of matter from their 
main-sequence companions \citep{1974SvA....18..217B, 1982Natur.300..728A}.

Based on the type of the companion star, binary MSPs are divided into several subclasses. Those MSPs bounded 
in tight orbits (binary periods P$_b < $ 20 h) with very low-mass companions (M$_c < 0.05~$M$_{\sun}$) are dubbed black widow (BW) pulsars. 
In these systems, the companion star is ablated by
the pulsar high-energy radiation and the wind of relativistic particles   until eventually may become fully evaporated. 
 Isolated MSPs that are not associated with globular clusters are believed to be formed in this scenario \citep{1988Natur.334..227V}. They 
currently account about 20 per cent of the total MSP population\footnote{http://astro.phys.wvu.edu/GalacticMSPs/}.

The first BW system, PSR 1957+20, was introduced by \citet{1988Natur.333..237F}. It is a MSP binary with P$_{b}$ = 9.2 h and the companion mass 
M$_c < 0.05$~M$_{\sun}$. Following this discovery, two more BW systems in the Galactic disk, PSR J2051$-$0827 and PSR~J0610$-$2100, were detected in the \textit{Parkes} 
radio surveys \citep{1996ApJ...473L.119S, 2006MNRAS.368..283B}. 
However, no significant progress in the field was made until the launch of the \textit{Fermi} Large Area Telescope. The \textit{Fermi} detections and 
radio follow-up searches have significantly expanded the general population of observed MSPs\footnote{https://fermi.gsfc.nasa.gov/science/eteu/pulsars/}, 
and, as a particular contribution, they increased the amount of known Galactic disk BWs 
from three to more than twenty\footnote{www.atnf.csiro.au/people/pulsar/psrcat/, \citet{2005AJ....129.1993M}} (see, e.g., \citet{ray, 
2016ApJ...819...34C, 2017ApJ...846L..20B} and references therein). 

\begin{table*}
\caption{Parameters of PSR J2052+1219. }
\label{t:656_prop}
\begin{tabular}{cccccccccccc}
\hline\hline
\multicolumn{6}{c}{Observed}&&\multicolumn{5}{c}{Derived} \\
\cline{1-8}\cline{10-12}
$P$        & $\dot{P}$   & $D\!M$       & $l$   & $b$ & RA  & DEC & $P_\mathrm{b}$          && $\tau$ & $B$                   & $\dot{E}$                  \\
ms         & 10$^{-21}$ & cm$^{-3}$ pc & ($\degr$)   & ($\degr$) &  (J2000)   &  (J2000)  & h   && Gyr    & G                     & erg~s$^{-1}$                \\
\hline
1.99    & 6.7      & 42         & 59 & -20 & 20:52:47.77844(5)        &12:19:59.0281(9)  &2.75 && 4.7   & 1.2 $\times$ 10$^{8}$  & 3.34$\times$ 10$^{34}$     \\
\hline \hline
\multicolumn{2}{c}{Orbital Period (d)} & \multicolumn{4}{c}{semi-major axis (lt-s)}&\multicolumn{4}{c}{mass function (10$^{-6}M_{\sun}$)} && Reference 
\\ \hline
\multicolumn{2}{c}{0.155} & \multicolumn{4}{c}{0.061}&\multicolumn{5}{c}{18.43} & \citet{2019arXiv190709778G} \\
\hline
\end{tabular}
\label{t:basic}
\end{table*}

These discoveries, in turn, opened up the possibility for multiwavelength studies of BW systems including the optical range. 
However, to date only a small fraction of new BWs  has been studied  in the optical \citep{2015salt.confE..75W}.  
Photometric information was reported for about a dozen of known Galactic disk BWs and only two systems, PSR~J1311$-$3430 
and PSR J1301+0833, were studied using spectroscopy \citep{2012ApJ...760L..36R, 2015ApJ...804..115R, 2016ApJ...833..138R}.  
Optical observations of BWs  are important to track their
evolution and to study the evaporation process and formation of isolated MSPs.
In addition, they allow us to set independent constraints on fundamental parameters of  these binary systems. 

\begin{figure*}
\setlength{\unitlength}{1mm}
\begin{center}
\begin{picture}(150,85)(0,0)

\put (76,0) {\includegraphics[width=85mm, bb =-10 90 580 665, clip=]{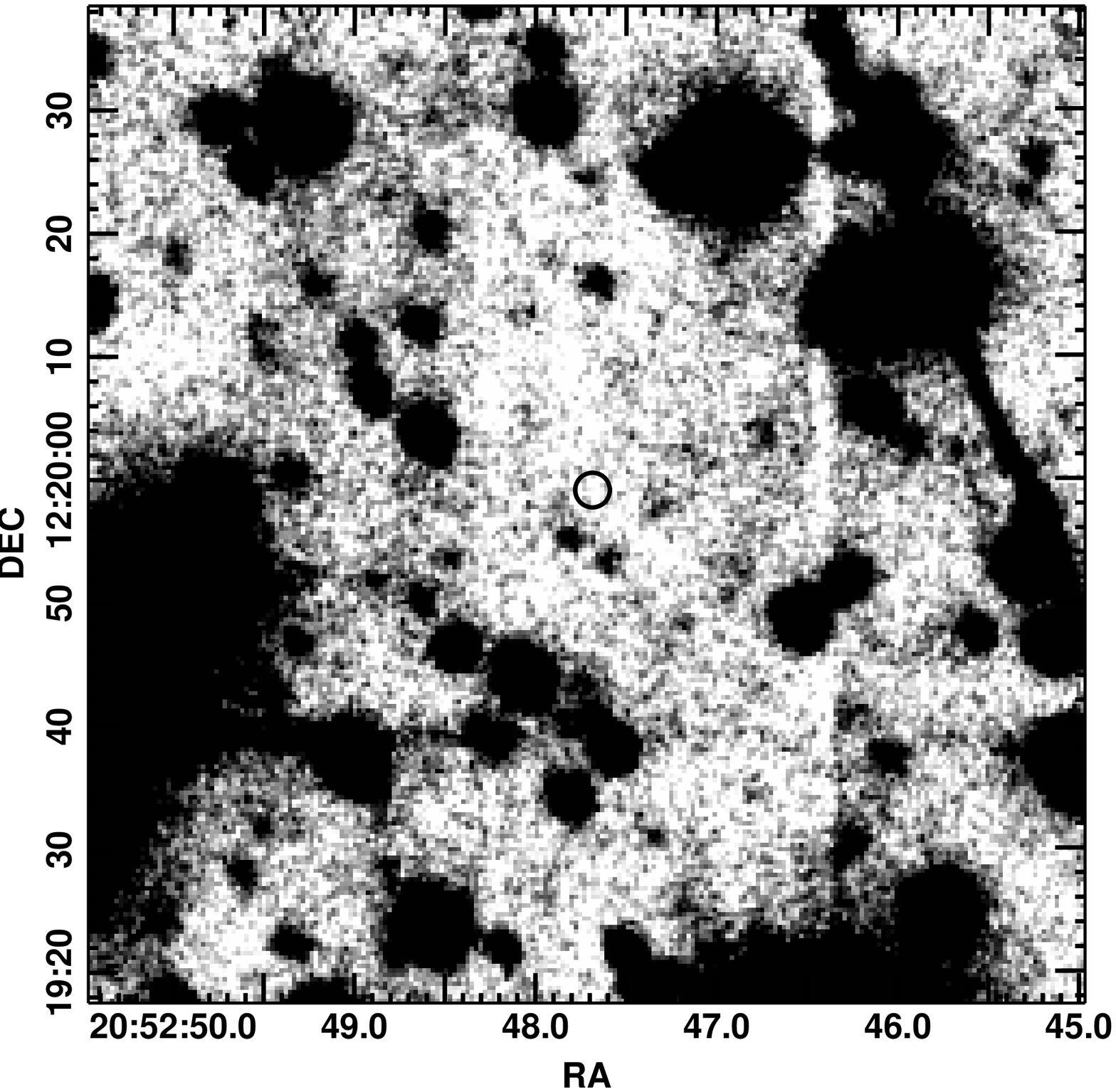}}
\put (-15,0) {\includegraphics[width=85mm, bb =-10 90 580 665, clip=]{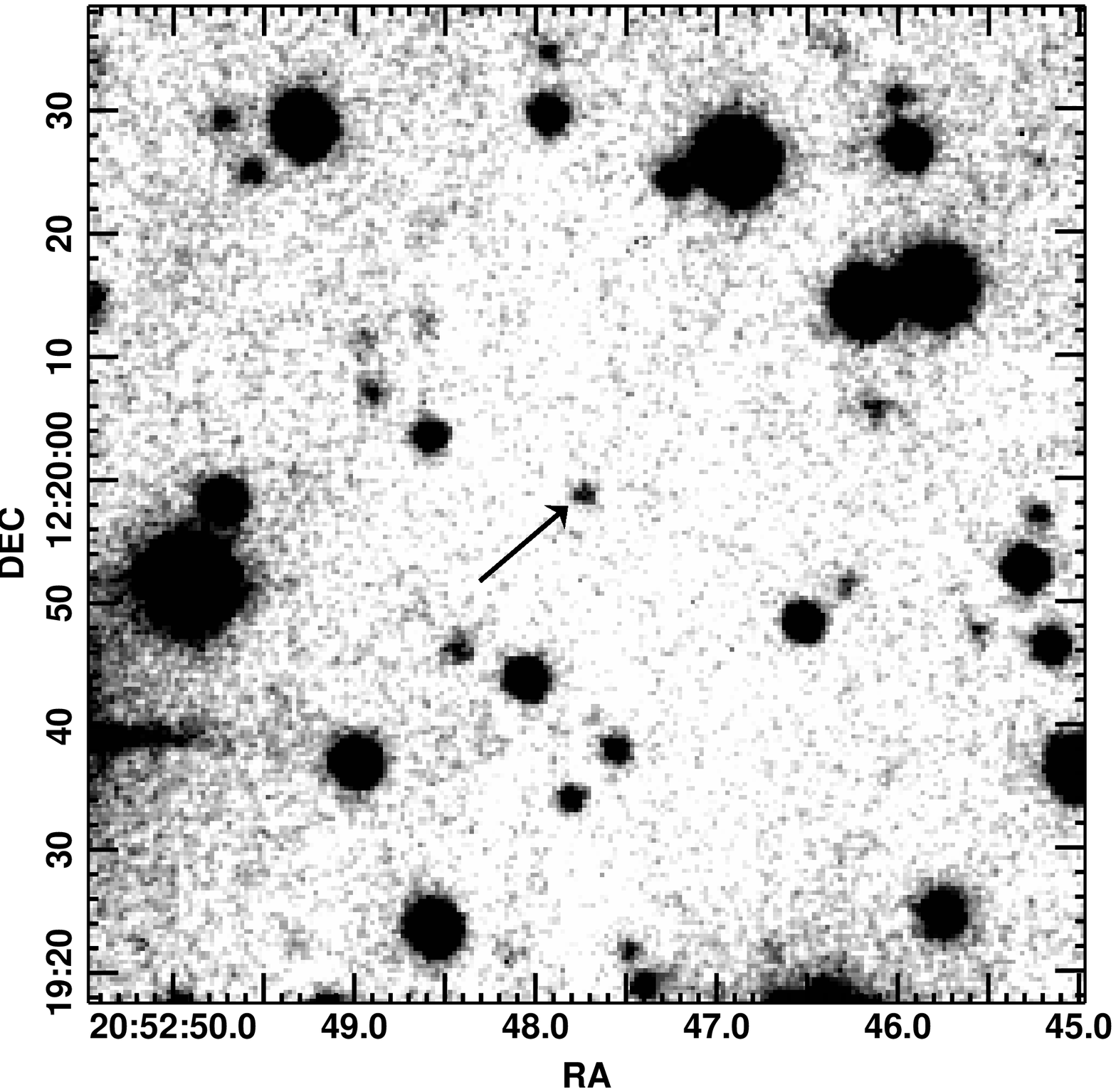}}
\put (45.5,62) {\includegraphics[width=22mm]{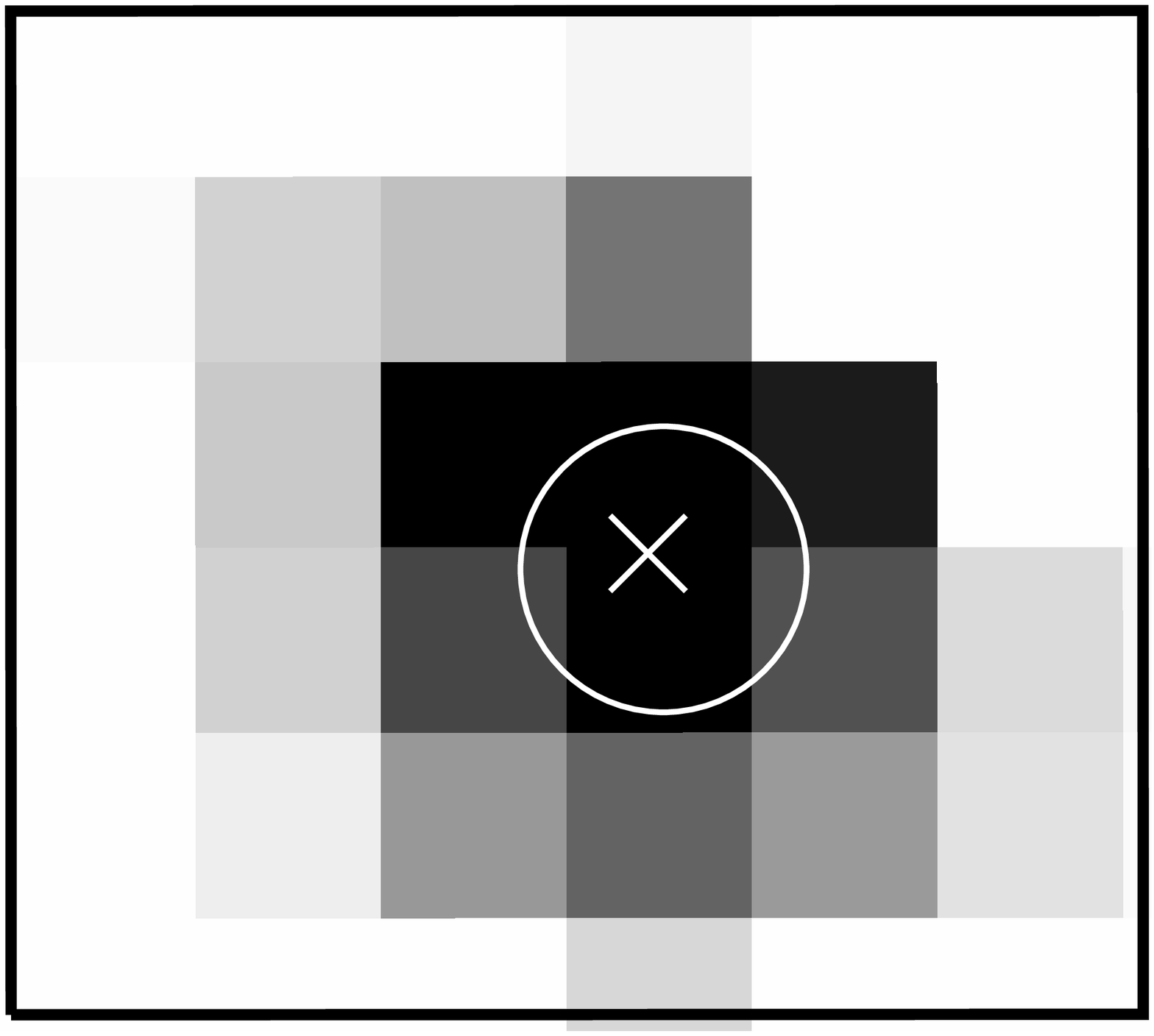}}
\put (10,30){\Large A}
\put (17,17){\Large B}
\put (6,77){\Large C}
\put (62,28){\Large D}
\end{picture}
\end{center}
\caption{
{\sl Left panel:} 
 $\approx$ 1.3\arcmin$\times$1.3\arcmin\ image
of the \psr~field obtained in the $R$-band.
The arrow points to the variable source located at the pulsar radio position. 
It is enlarged in the top-right corner. The cross 
shows its centre on the image and the circle corresponds to the 3$\sigma$ pulsar radio timing position uncertainty.
The letters mark the secondary photometric  standards  in the pulsar field from Table.~\ref{tab:SecSts}.
The image corresponds to the object's peak brightness. 
{\sl Right panel:} Deeper  image 
of the same field in the same band obtained near the minimum brightness stage of the same 
source when it falls below the detection limit. The source position is marked by the circle.}
\label{fig1}
\end{figure*}

The millisecond PSR~J2052$+$1218 was recently discovered in the Arecibo telescope search of unidentified
gamma-ray sources from the \textit{Fermi} Large Area Telescope (LAT) four year point source catalogue \citep{2016ApJ...819...34C}. Its 
coordinates obtained from subsequent radio timing measurements are RA$_{2000}$ = 20:52:47.77844(5) 
and Dec$_{2000}$ = 12:19:59.0281(9). We will hereafter refer to this pulsar as PSR J2052+1219. 
The pulsar spin period is $P$=1.99 ms with $\dot{P} = 6.7\times 10^{-21}$, its 
characteristic age is $\tau=4.7$~Gyr, and the spin-down luminosity is $\dot{E}$ = $3.34\times 10^{34}$ erg~s$^{-1}$ 
for a NS moment of inertia of $I=10^{45}$ g cm$^2$. It is found  
in an eclipsing binary system with $P_{b}$ $\approx$ 2.8~h 
containing a very low-mass companion (Table~\ref{t:basic}).  
Using Keplarian parameters derived from orbital
timing solution, \citet{2016ApJ...819...34C} have calculated a minimum companion mass of $\gtrsim$0.033 M$_{\sun}$ suggesting the system is a BW. 
Based on the dispersion measure DM=42 pc cm$^{-3}$ and the NE2001
Galactic electron-density model \citep{cordes}, 
they estimated a distance to PSR J2052+1219 of 2.4 kpc.
However, the pulsar distance based on the YMW16 electron-density model \citep{2017ApJ...835...29Y} is about 3.92 kpc. 
Accounting for a large  Galactic latitude of the pulsar, 
$b\approx -20\degr$, at such distances it has to be located about 1 kpc 
above the Galactic disk. This implies a low interstellar 
extinction in its direction. Indeed, according to the 
Galactic 3D extinction/distance maps\footnote{http://argonaut.skymaps.info}, 
the reddening along its line 
of sight reaches an upper limit of  
$E(B-V)=0.12^{+0.02}_{-0.02}$ and remains constant at 
distances $\ga$1.5~kpc.  
The older measurements\footnote{https://irsa.ipac.caltech.edu/applications/DUST/} give slightly smaller values of the total Galactic absorption in the pulsar direction: $E(B-V) = 0.0844\pm0.0053$ \citep{2011ApJ...737..103S} and $E(B-V) = 0.0981\pm0.0062$ \citep{1998ApJ...500..525S}. Therefore, taking into account the uncertainties,
below we consider the colour excess at the pulsar distance to be in a range of $E(B-V) = 0.08-0.14$.

Inspection of the Panoramic Survey Telescope and Rapid Response System Survey (Pan-STARRS; \citet{flewelling}) catalogue 
allowed us to reveal 
a possible optical counterpart of the pulsar companion,
PSO J205247.782+121959.174, with $r'\approx 22.3$. 
To confirm it and determine the physical parameters of the source, we performed optical time-resolved multi-band observations.

\section{Observations and data reduction}

\begin{table}
\caption{Log of the PSR J2052+1219 observations with the 2.1m telescope at the OAN-SPM.}
\label{log}
\begin{tabular}{ccccccccc}\hline
Date & Filter& Exposure time, & Airmass & Seeing, \\ 
 & & seconds & & arcsec \\
\hline
14/09/2018 & $R$ & 600x38 & 1.1$-$2.2 & 1.4--2.2 \\
15/09/2018 & $R$ & 600x34 & 1.1$-$1.7 & 1.5--2.4\\
16/09/2018 & $V$ & 600x16 & 1.1$-$1.2 & 1.5--1.8 \\
 & $R$ & 600x16 & 1.1$-$1.6 & 1.4--1.7\\
17/09/2018 & $B$ & 600x2 & 1.4$-$1.5 & 1.5--1.6\\
 & $V$ & 600x2 & 1.4$-$1.7 & 1.4--1.5 \\
 & $R$ & 600x28 & 1.1$-$1.8 & 1.5--1.6\\
 & $I$ & 600x2 & 1.5$-$1.9 & 1.3--1.5\\
\hline
\end{tabular}
\end{table}

The time-resolved photometric observations of the pulsar field were performed with the "Rueda Italiana" instrument\footnote{www.astrosen.unam.mx} 
attached to the 2.1m telescope at the Observatorio Astronomico Nacional San Pedro M\'artir (OAN-SPM), Mexico 
 on September 14$-$17, 2018. The field of view of the detector was $6\arcmin\times6\arcmin$ with an image scale of $0\farcs34$ in the 2$\times$2 CCD pixel binning mode. 
 The conditions during the observing run were clear. 
The observing log is presented in Table~\ref{log}. 
\begin{figure*}
\setlength{\unitlength}{1mm}
\resizebox{15.cm}{!}{
\begin{picture}(130,130)(0,0)
\put (-2,87){{\includegraphics[width=13.95cm,  bb=55  50 715 250, clip=]{zharikovFig2a.eps}}}
\put (36,77)  {{\includegraphics[width=0.95cm,  bb=284 340 327 384, clip=]{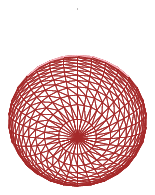}}}
\put (52,77)  {{\includegraphics[width=1.2cm,  bb=427 370 483 412, clip=]{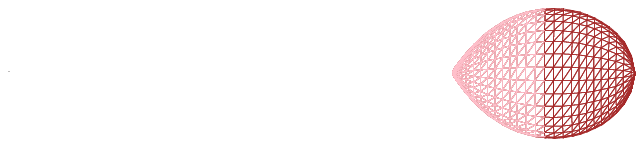}}}
\put (66,77){{\includegraphics[width=0.95cm,  bb=283 402 327 446, clip=]{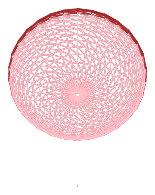}}}
\put (76,77)  {{\includegraphics[width=0.95cm,  bb=193 397 238 437, clip=]{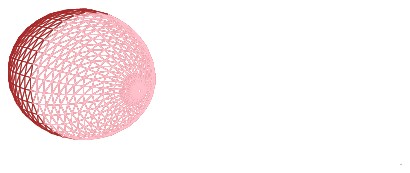}}}

\put (-8,0){{\includegraphics[width=14.6cm,    bb=25  40 715 470, clip=]{zharikovFig2b.eps}}}

\end{picture}}
\caption{Upper panel: Power spectrum of the variable source based on the $R$-band data.  
The main peak corresponds to an orbital period of 2.752 h. 
Lower panel, top: Observed $BVRI$ light curves of the source folded 
with 
the orbital period  and 
best fits to the data 
 (solid lines) using the model described in the text. 
 Two periods corresponding to the orbital phase range -0.5---1.5 
 are shown for clarity. The model shapes  of the secondary and the brightness distribution over its surface   as visible by an observer 
 at phases 0.0, 0.25, 1.0 and 1.125 are  
 shown at the top of the plot. Lower panel, bottom: The observed minus calculated (O-C) light curves.}
\label{fig2}
\end{figure*}
\begin{table}
\caption{Secondary photometric standards  marked in Fig.~\ref{fig1}.}
\label{tab:SecSts}
\begin{tabular}{lccccc}
\hline
Star  & B&V &R& I \\
A&20.95(1) & 19.71(1) & 18.97(2) & 18.43(2) \\
B&19.65(1) & 18.95(1) & 18.52(1) &18.12(1) \\
C&19.40(1)  & 18.40(1) &17.80(1) & 17.23(1) \\
D& 19.51(1) & 18.55(1) &17.95(1) & 17.31(1) \\  \hline 
\end{tabular}
\end{table}

We carried out standard data processing, including bias subtraction and flat-fielding with the Image Reduction and Analysis Facility ({\tt IRAF}) package. The cosmic rays 
were  removed from all images using the L.A.Cosmic algorithm \citep{2001PASP..113.1420V}. 
 Astrometric referencing  was performed on a single 600s exposure obtained in the best-seeing conditions. We used a set of 9 stars from 
the Gaia DR2 Catalogue \citep{2016A&A...595A...1G, 2018A&A...616A...1G} with  positional errors of $\la$0.13 mas. 
Formal $rms$
uncertainties of the resulting astrometric fit were $\Delta$RA~$\la$ $0\farcs09$ 
and $\Delta$Dec~$\la$ $0\farcs07$. The catalogue conservative uncertainty of 0.7 mas \citep{2018A&A...616A...2L} can be neglected in case of our astrometric solution. 

The photometric calibration was performed using reference stars from the 
PG2213$-$006 
standard field \citep{1992AJ....104..340L} observed on September 17, 2018 immediately after the target. 
Using their instrumental magnitudes and the site extinction coefficients $k_{B}$=0.25, $k_{V}$=0.14, $k_{R}$=0.07 and $k_{I}$=0.06 \citep{sp,spg}, 
we calculated the zero-points for this night: $Z^L_B$=24.88$\pm0.01$, $Z^L_V$=25.15$\pm0.01$, $Z^L_R$=25.00$\pm0.01$ and $Z^L_I$=24.20$\pm0.01$. 
The zero-points were verified using $\sim$20 stars in the target field whose magnitudes were extracted  from
 the Pan-STARRS catalogue. The Pan-STARRS $gri$ 
 magnitudes \citep{2012ApJ...750...99T} were transformed to the $BVRI$ system using the equations (1---7)
 and coefficients from Table~2 of \citet{2018BlgAJ..28....3K}. Using the obtained values, we then calculated the zero-points $Z^P_B$=24.78$\pm0.01$, $Z^P_V$=25.05$\pm0.01$, $Z^P_R$=24.83$\pm0.01$ and $Z^P_I$=24.13$\pm0.01$. We found a "grey" shift of  $\approx$ 0.10   between the $Z^L$ and $Z^{P}$ values. Revising several field stars on different time-resolved images, we 
 found a slight transparency variation of about 0.1 mag. Therefore, in the following analyses 
 we only use the $Z^{P}$ values.

 To calibrate the data obtained during the other nights,  
we measured magnitudes of four stars in the pulsar vicinity, which we  defined as secondary photometric standards. 
They  are marked in the left panel of Fig.~\ref{fig1} with $A$, $B$, $C$ and $D$, and 
their magnitudes are given in Table~\ref{tab:SecSts}.

 \begin{figure*}
\setlength{\unitlength}{1mm}
\resizebox{15.cm}{!}{
\begin{picture}(130,145)(0,0)
\put (-14,0){{\includegraphics[width=14.950cm,bb = 20 120 590 660, angle=0, clip=]{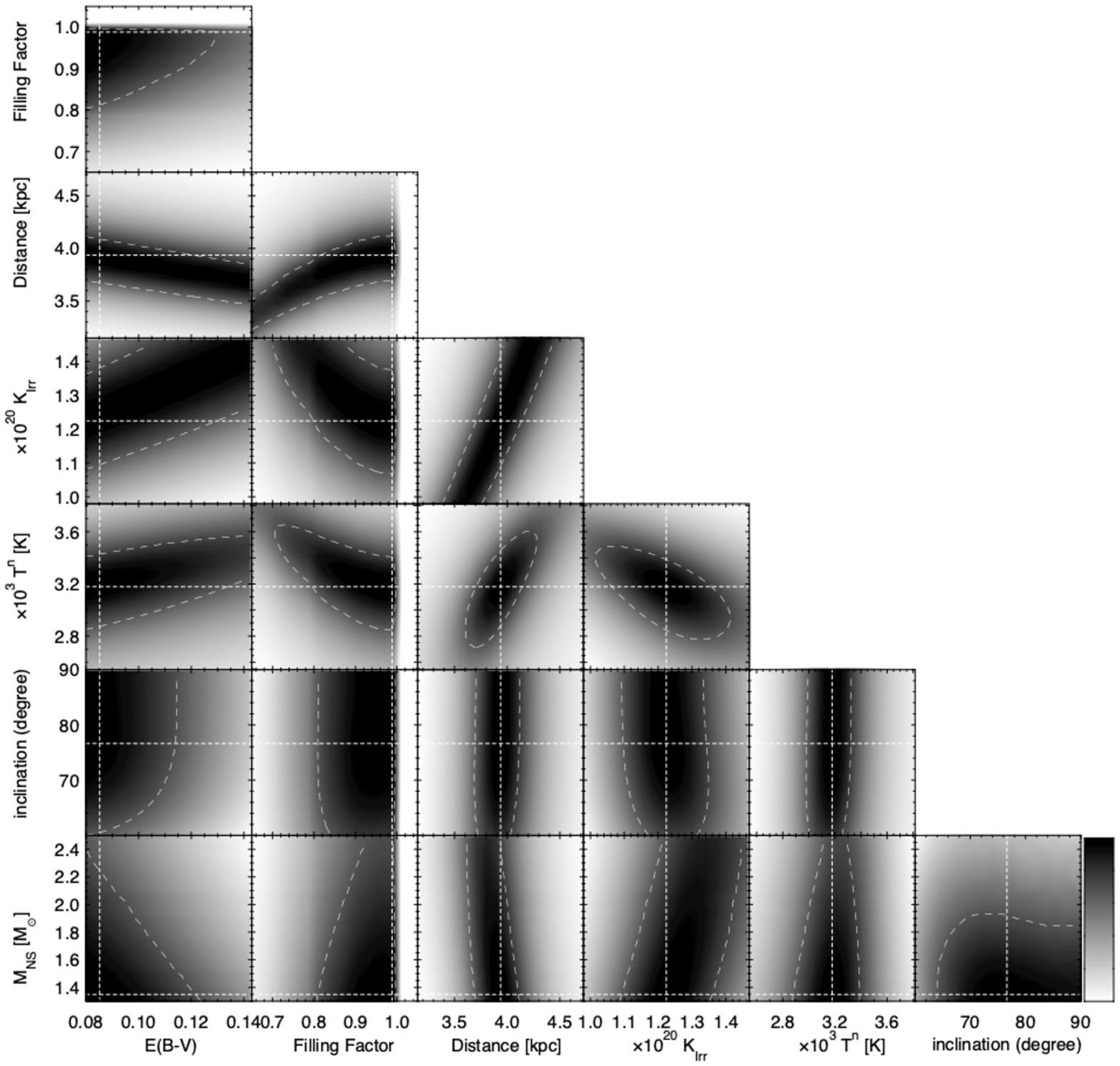}}}
\put(136,10){0.0}
\put(136,32){1.0}
\put(90,118){$M_{NS} = 1.35^{+0.5}_{-0.05}M_\odot $}
\put(90,113){$M_{2} = 0.034\pm0.011 M_\odot $}
\put(90,107){$i = 77\degr\pm 13 $}
\put(90,102){Filling factor = $1.0^{+0.00}_{-0.17}$}
\put(90,97){$E(B-V)=0.085^{+0.030}_{-0.005}$}
\put(90,92){D(kpc)$ = 3.94\pm0.16$}
\put(90,87){$K_\mathrm{irr}  = 1.22(\pm0.7)\times 10^{20}$}
\put(90,82){$T^n_2(K) = 3200\pm200$}
\put(90,77){$R_2 (R_{\sun}) \la 0.15$}
\end{picture}}
\caption{Errors of the fit. 
The black-and-white  scale corresponds to the 
minimum (black) and maximum (white) values of $\chi^2=f(par1, par2)$ in the corresponding plot when other parameters 
are fixed at the best values. The short-dashed lines correspond to the global minimum of all fitted parameters. The irradiation factor $K_\mathrm{irr}$ is given in $\mathrm {ergs~cm^{-2}~s^{-1}~sr^{-1}}$.
The thin long-dashed lines 
show the $1\sigma$ errors of the fit parameters. }
\label{fig3}
\end{figure*}

\section{Results}
\label{Results}

The $R$-band images 
of the pulsar field are 
presented on both panels of Fig.~\ref{fig1}. The arrow on the left panel points towards a source which shows strong variability.
It is clearly visible on several subsequent images and is absent on others 
 obtained during the same night. 
The $\sim 2\arcsec \times 2\arcsec$ vicinity of the object is enlarged in the top-right corner of the left panel, and 
its  peak position 
is shown by the white cross.
The centre of the circle corresponds to the pulsar radio position.
The circle radius of 0.27$\arcsec$ represents the 3$\sigma$ pulsar radio position uncertainty 
that accounts for the optical astrometric referencing and the radio timing uncertainties.

\begin{table}
\caption{System parameters used for the $BVRI$-band light curve modelling.}
\label{model}
\begin{tabular}{llll}
\hline \hline
\multicolumn{4}{l}{ Variable and their allowed ranges:} \\ \hline
\multicolumn{4}{c}{$1.3<M_{NS}<2.5$ } \\ 

\multicolumn{4}{c}{$0.08<E(B-V)<0.14$ } \\ 
\multicolumn{4}{c}{ R mag (phase=0.0) $\ga$ 24.7} \\
\multicolumn{4}{c}{ T$^n_2$; Roche Lobe Filling Factor;   Distance} \\
 \multicolumn{4}{c}{ Irradiation Factor $K_\mathrm{irr}$, system inclination $i$} \\  \hline
{ Result:}    & & & \\ \hline
\multicolumn{4}{c}{$M_{NS} = 1.35^{+0.5}_{-0.05}M_\odot$} \\
\multicolumn{4}{c}{$M_{2} = 0.034\pm0.011 M_\odot $} \\
\multicolumn{4}{c}{ $E(B-V)=0.085^{+0.030}_{-0.005}$}\\
\multicolumn{4}{c}{  $T^{n}_2$=3200$\pm$200  K}\\
\multicolumn{4}{c}{0.12 < $R_2$/$R_{\sun}$  $\lesssim$ 0.15 }\\
\multicolumn{4}{c}{0.83 < Roche Lobe Filling Factor  $\lesssim$ 1.0 }\\
\multicolumn{4}{c}{D (kpc)=3.94$\pm$0.16 }\\
\multicolumn{4}{c}{$K_\mathrm{irr} \mathrm{(erg~cm^{-2}~s^{-1}~sr^{-1}}) =1.22(\pm0.7)\times 10^{20}$ }\\
\multicolumn{4}{c}{ $i$=77\fdg0$\pm13\fdg0$ } \\
\multicolumn{4}{c}{  $T^{d,min}_2$=3630 K < $T^{d}_2$ <  $T^{d,max}_2$=6530 K} (derived)\\

 \hline 

\end{tabular}
\end{table}
The 
source and the pulsar positions coincide perfectly, 
suggesting their  association. 
We performed time-resolved aperture photometry of the source. 
The object magnitude at the maximum is $R_c\approx22.0$, whereas on some other images the brightness of the source falls below the $3\sigma$ detection 
limit\footnote{ The detection limits presented in this article were derived following the standard procedure (see, e.g., \citet{2013MNRAS.435.2227Z}).}   
$R_c^\mathrm{lim.}\sim24.1$ of a single 600~s exposure. 

\begin{table*}
\begin{center}
\caption{Parameters of BW systems detected in the optical. The first column corresponds to the name of the object, 
$\Delta m$ is the full amplitude of the optical light curve variation in the corresponding band, $P_{orb}$ is the orbital period, 
$\dot{E}$ is the spin-down luminosity, $i$ is the system inclination, and the last four columns correspond to the "day-side" and 
"night-side" temperature estimations, the radius of the companion star and the respective references.}
\label{MSPs}
\begin{tabular}{l|ccc|cccc|c}
\hline
MSP                            & $\Delta$m                                   &  P$_{orb}$    &$\times$10$^{34}$  $\dot{E}$  & inclination       & $\times$10$^3$ $T_2^{d}$ &$\times$10$^3$ $T_2^{n}$& R                          & Ref.  \\ 
                                    &                                                    &  (h)               & (ergs s$^{-1}$)                                    & degree            & (K)                                       &   (K)                                   & (R$_{\sun}$ )        &          \\ \hline
 J1311$-$3430             &                                                   &  1.56             & 4.9                                          &                        &                                              &                                          &                              & 13        \\
J0636+5128                 &  $\Delta g \approx$2.0               &  1.6               & 0.58                                        & 24(2), 40(6)    &4.7                                        & 1.7                                     &    0.08                   & 4, 5       \\
                                     &                                                   &                      &                                                 &                        & 3.9                                      & 2.5                                     &  0.10                     &              \\
 J1518$+$0204C$^{*}$         &  $\Delta F{606W} > 1.3$            & 2.1                & 6.7                                         &                        & 3.4$-$5.3                              &                                          &                              & 12        \\
 J2051$-$0827            & $\Delta F{675W}\approx 3.3$     & 2.38              & 0.55                                         & $\sim$40         & 4$-$4.7                               & $< 3$                                 &                              & 10,15  \\
 \textbf{J2052+1219}   & $\Delta R >$2.7                           & 2.8                &        3.34                                         & $77(13)$   & 6.6                                         & 3.5                                     & 0.14                        & 2         \\ 
 &  $\Delta r\gtrsim4.4$&&&&&&& 16\\
 J1544+4937               &  $\Delta g \approx$2.0                 &  2.9              &1.2                                           & $\approx$60   &5.4                                         &  3.9                                   & 0.037                    & 7          \\
  J1810+1744              &  $\Delta g \approx$4.0                 & 3.6               & 3.97                                        & 48(7)               &10.0                                       & 3.1                                    & 0.15                      &1,3        \\
 J1953+1846A$^{*}$             &  $\Delta F{606W}\approx 3.3$     & 4.2                &  1.64                                              & $\sim$90         &                                             &                                           &                              & 8        \\
 J0023+0923               & $\Delta g \approx 4.6$                 & 4.8               &1.51                                          & 58(14)            &4.8                                       & 2.9                                     & 0.05                      & 1           \\
 J2256$-$1024            & $\Delta g \approx$5.8                  &  5.1               & 3.95                                        &  68(11)            &4.2                                        & 2.5                                    &0.05                        & 1        \\
  J0952$-$0607           & $\Delta r \approx$1.6                   &  6.42             & $<$16.0                                  &    $\sim$45     &              4.5-5.8                    &   $\sim$2.5                                       &                              & 9         \\ 
 J1301+0833               &$\Delta g > $2.5                            &  6.5               & 7.0                                          & $\approx$52   &4.6                                        & 2.7                                    &  $<$0.1                   & 6         \\
  J0610$-$2100           & $\Delta R > $1.7                          &  6.86             & 0.23                                        &                        & 3.5                                      &                                           &                              &  14        \\
  B1957$+$20              & $ \Delta R \approx $ 5.1                &  9.17             & 16                                           & 65(2)               & 2.9                                       & 8.3                                     &                              & 11        \\

 \hline
\end{tabular} \\
\begin{tabular}{l}
1-\citet{2013ApJ...769..108B},
2-this paper, 
3-\citet{2014ApJ...793...78S}, 
4-\citet{2018ApJ...864...15K}, 
5-\citet{2018ApJ...862L...6D} \\
6-\citet{2016ApJ...833..138R}, 
7-\citet{2014ApJ...791L...5T}, 
8-\citet{2015ApJ...807...91C},
9-\citet{2017ApJ...846L..20B},
10-\citet{2001ApJ...548L.183S},\\
11-\citet{2007MNRAS.379.1117R},
12-\citet{2014ApJ...795...29P},
13-\citet{2015ApJ...804..115R},
14-\citet{2012ApJ...755..180P},
15-\citet{1996ApJ...473L.119S} \\
16-\citet{2019arXiv190800992D} \\
$\dot{E} = 4\upi^2I(\dot{P}/P^3)$, for  a moment of inertia $I = 10^{45}$ g cm$^{-2}$, pulse period $P$, and period derivative $\dot{P}$ \\
$^{*}$ { The intrinsic $\dot{P}$ and corresponding $\dot{E}$   can be different
from the measured ones due to acceleration in the globular clusters M5 and M71.}
\end{tabular}
\end{center}
\end{table*}
The obtained $R$-band photometric data were analysed to search for periodicity using the Discrete Fourier Transform code \citep{1975Ap&SS..36..137D}.  The resulting power spectrum is presented on the top panel of Fig.~\ref{fig2}. The peak corresponding to the maximum power yields the photometric period of $P_{phot}$ = 2.752~h, 
which is in agreement with the pulsar binary period $P_{b}$ = 2.75~h (Thankful Cromartie, private communication). 
The  light curves folded  with the photometric period  are shown in the bottom panel of Fig.~\ref{fig2}.
The phase 0.0, defined as the time when the secondary is placed  between the pulsar  and an observer,  corresponds to HJD$_0$=2458374.74535.
The brightness variation in other bands is  consistent with that in the $R$-band.  
To detect the object at the minimum, we summed all the images obtained close to the phase 0.0. However, we have only obtained a $3\sigma$ upper limit on its minimum brightness of $R_\mathrm{min}\ga$24.7.
The respective image is shown in the right panel of Fig.~\ref{fig1}.

The coordinate accordance, the coincidence of the photometric and  binary periods, 
and the similarity of the 
light curve shape to those of other BWs (see Table~\ref{MSPs})  strongly
suggest that the detected variable source is the binary companion 
of PSR J2052+1219. 

The accepted  interstellar reddening  towards the pulsar   
$E(B-V)=0.08\div0.14$ (see Sect.~\ref{Intro})
gives Johnson-Cousins $BVRI$ extinction  values  
$A_B=0.29\div0.44$, $A_V=0.22\div0.33$, $A_R=0.17\div0.26$, and $A_I=0.12\div0.18$
 for $R_V=3.1$ \citep{2011ApJ...737..103S}.
 Therefore, the unabsorbed brightness of the source at the maximum is
$R_c\approx 21.8$. As we noted before, \citet{2016ApJ...819...34C} have calculated 
a minimum companion mass of $\ga$0.033 M$_{\sun}$ assuming the pulsar mass $M_\mathrm{NS}=1.35M_{\sun}$   and the orbit inclination angle $i=90$\degr.  
A corresponding mass function  $f(M_\mathrm{NS}, M_{2})$ of  18.78$\times10^{-6}M_{\sun}$ 
is slightly different from the updated value presented by  \citet{2019arXiv190709778G}   
(see Table~\ref{t:basic}), while its uncertainty and the uncertainty of the projected semi-major orbit axis still remain unknown.  In any case,  such a small difference, likely reflecting their uncertainties, does not affect the results of our light curve analysis presented below. 
For such a low-mass object, the most likely radius value should be about $\sim 0.1 R_{\sun}$ \citep{2009AIPC.1094..102C}. 
In these old binaries, the secondary is mainly responsible for the optical radiation. However, such a small object would be very faint and hardly  
detectable if isolated. This is consistent with the R-band upper limit at the phase 0.0. In close binary MSP systems, the irradiation from 
the pulsar increases the temperature and luminosity of 
the front side of the companion and makes it observable at respective orbital phases.   
For this reason, 
the strong variability of the detected source is explained by a high temperature difference between the front 
and the back sides of the secondary. 

 Unfortunately,
the available radio data do not provide information on the duration of 
the radio emission eclipse in an orbit.  Typically, MSPs in eclipsing binaries are eclipsed for 10--40 per cent of an orbit at 2 GHz, implying the presence of ionised material in a region larger than the Roche lobes of the companions \citep{2009Sci...324.1411A}.
Taking into account this fact and the system parameters, 
we reproduced the $BVRI$-band light curves using 
the modelling technique developed by \citet{2013A&A...549A..77Z}. It was designed to 
include different types of the primary, such as a black hole, a NS, a white dwarf or a main-sequence star, the filling factor of the Roche lobe  
 by the secondary, and possible accretion structures in 
a binary system. The irradiation of the secondary was 
taken into account as 

\begin{equation}
     T^{d}_2 = T^n_2*\left(1 + \frac{F_\mathrm{in}}{\Delta S\sigma (T^n_2)^4 }\right)^{1/4},
\end{equation}
where $T^{d}_2$ 
and $T^n_2$ are the temperatures 
of the "day-side" and "night-side"
surface element of the secondary, respectively,
$F_\mathrm{in}$ is the effective heating flux incoming to the "day-side" surface element $\Delta S$  of  the secondary, and $\sigma$ is the Stefan-Boltzmann constant. The effective heating in our model is defined by the irradiation factor $K_\mathrm{irr}$  [$\mathrm{ergs~s}^{-1}~\mathrm{cm}^{-2}~\mathrm{sr}^{-1}$] 
as 

\begin{equation}
F_\mathrm{in} = \cos(\alpha_\mathrm{{norm}}) ~ \Omega ~ \Delta S ~ K_\mathrm{{irr}}
\end{equation}

where $\alpha_\mathrm{norm}$ is the angle between the incoming wind/flux
and the normal to the surface element $\Delta S$, 
$\Omega = \upi R_\mathrm{NS}^2/a^2 \approx 1\times 10^{-9}$ is the solid angle from which the pulsar is visible from the 
surface element of the companion.
 The factor $K_\mathrm{irr}$ can generally  take into account  the heating effects caused by a combination of 
the pulsar thermal emission from the surface of the NS $\propto \sigma T_\mathrm{NS}^{4}$, its nonthermal radiation of the 
magnetospheric origin, and the pulsar wind. The last two are proportional to the pulsar spin-down luminosity $\dot{E}$.   
The factor defines 
the distribution  of the temperature on the "day-side" surface elements of the secondary $T_2^d$.
The MSPs are very old and cold NSs with $T_\mathrm{NS}\la 10^5$ and contribution of 
the thermal emission into the heating can be neglected.  
In this case, assuming the isotropy of the pulsar wind/radiation,  $K_\mathrm{irr}$   
can be expressed in terms of the measured  $\dot{E}$  and 
the heating efficiency $\eta$   
of the secondary 
defined  as
\begin{equation}
  \eta= \frac{\sum \Delta S \sigma (T_d^4 - T_n^4)}{\dot{E}\upi R^2/4\upi a^2},
\end{equation}
  where $a$ is the orbital separation, 
  $R$ is 
  the polar radius of
  the secondary and the  summation is performed over all surface elements $\Delta S$. 
  Combining equations (1 -- 3) yields 
 \begin{equation}
 K_{irr} =  \frac{ \eta \dot{E} }{4\upi^2 R_{NS}^2}, 
 \end{equation}
 assuming that $R$ is roughly equal to the radius of the secondary neglecting its Roche lobe filling. 
 In case of \psr, $\frac{  \dot{E} }{4\upi^2 R_{NS}^2} = 5.9\times 10^{20}$
$\mathrm{ergs~s}^{-1}~\mathrm{cm}^{-2}~\mathrm{sr}^{-1}$ 
for R$_\mathrm{NS}$ = 12~km.

To fit the data, 
we fixed the  mass function at the updated  value presented in Table~\ref{t:basic}: 
\begin{equation}
\label{massfunc}
 f(M_{NS}, M_2) = \frac{(M_2~sin~i)^3}{(M_{NS}+M_2)^2} =  0.00001843M_{\sun} 
\end{equation}

\begin{figure}
\setlength{\unitlength}{1mm}
\resizebox{15.cm}{!}{
\begin{picture}(130,55)(0,0)
\put (0,0){{\includegraphics[width=7.5cm, clip=]{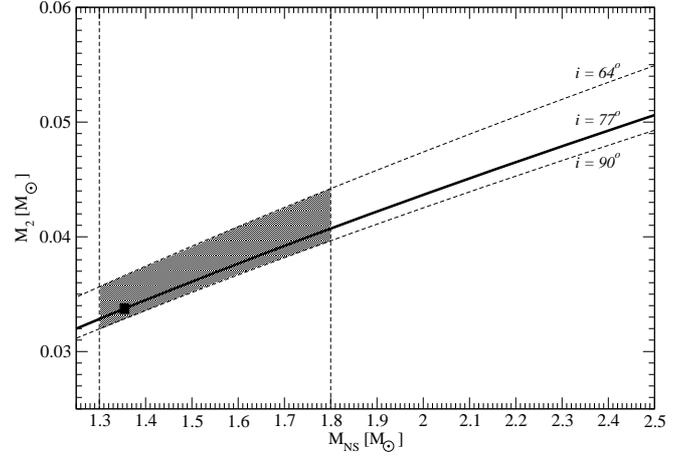}}}
\end{picture}}
\caption{Mass of the companion vs. mass of the pulsar. The filled area shows the 1$\sigma$ error box of the system parameters. The black square corresponds to the best values of the fit.}
\label{fig:MassFunc}
\end{figure}
 The variable parameters of the fit were  the mass of the pulsar M$_{NS}$, the distance, the "night-side" temperature of the secondary $T_2^n$,
 the system  inclination~$i$, 
the radius  $R_2$ or the Roche lobe filling factor of the secondary, the interstellar absorption $E(B-V)$, and the effective irradiation factor $K_\mathrm{irr}$.
The mass of the secondary M$_2$ and the mass ratio $q\equiv M_2/M_{NS}$ were determined from the equation~\ref{massfunc}.

 The  gradient descent method was used to find the minimum of the $\chi^2$ function
defined as
\begin{equation}
 \chi^2 = \sum^{B,V,R,I}_j  \sum^{N_k}_k \frac{(\mathrm{mag}^\mathrm{obs}_k - \mathrm{mag}^\mathrm{calc}_k)^2}{(\Delta \mathrm{mag}^\mathrm{obs}_k)^2}
\end{equation}

with the additional condition that the source is not detected at the phase 0.0 down to $R\approx 24.7$. $N_k$ is the number of the binary phase bins. For each of the observed phases ($k$)  the model magnitudes  $\mathrm{mag}^\mathrm{calc}_\mathrm{k}$ in each band were calculated from the integrated 
flux $\sum  \Delta S_\mathrm{n} \times   R_\mathrm{tr}(\lambda) \times BB_{\lambda}(T)$ of all visible elements of the system, where $R_\mathrm{tr}$ is the filter transmission, $\Delta S_\mathrm{n}$ is the projection of the surface element area to the line of sight, $BB(T)$ is the black body flux from the element. 
The total flux was converted into the magnitude $\mathrm{mag}^\mathrm{calc}_k$ taking into account  the distance to the system, the interstellar extinction, and the band's zero-points.
As a first step, the error of the fitting was selected arbitrarily. After the  minimum of the functional was reached, 
we searched for the global minimum several times decreasing the acceptable fitting error in the vicinity of the minimum  
that was reached in the previous step. The search was repeated until the difference between the model and observational light curves became insignificant.

The result of our light curve fitting is presented in Figs.~\ref{fig2},~\ref{fig3},~\ref{fig:MassFunc}, and  Table~\ref{model}.
As  seen from Fig.~\ref{fig2}, the observed  light curves are perfectly fitted by the model
although the resulting formal  reduced $\chi^2/DOF$ appears to be  relatively large (3.54)  and 
the differences between the observed and calculated (O-C) magnitude points reach  $\pm0.3$ mag. Because the measured magnitude errors are much smaller,  the dispersion of the observed points around the model  curves 
is likely caused by an intrinsic stochastic  variability of the source at smaller time scales than $P_\mathrm{b}$, which is not accounted by the model.   

\begin{figure*}
\setlength{\unitlength}{1mm}
\resizebox{15.cm}{!}{
\begin{picture}(130,77)(0,0)
\put (-3,0){{\includegraphics[width=13.0cm, clip=]{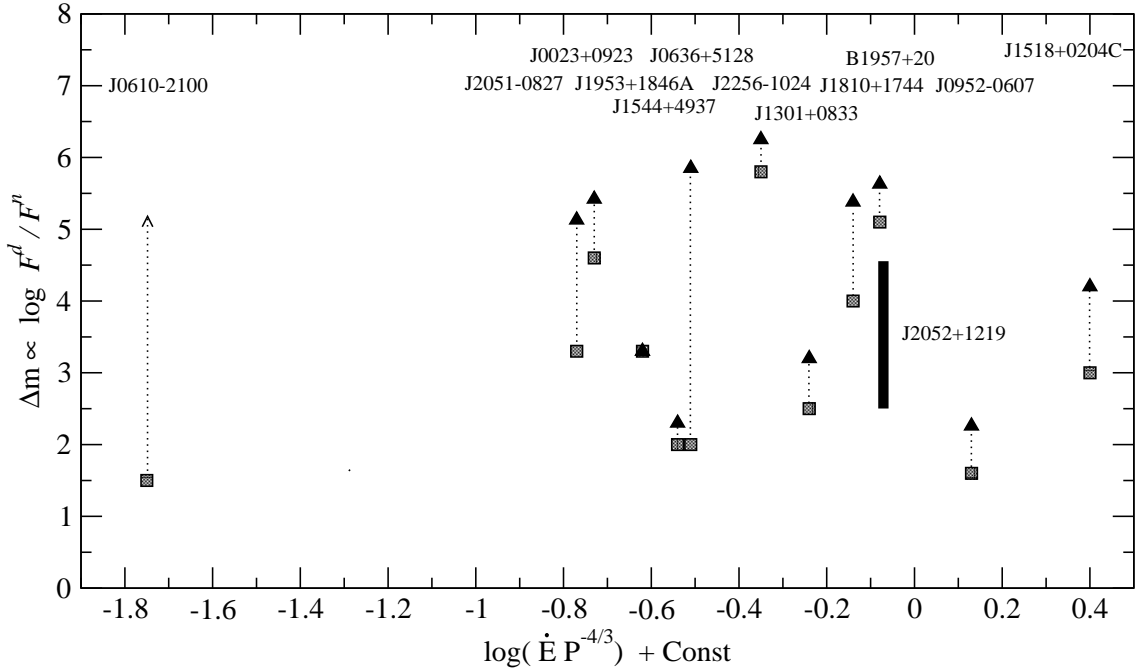}}}
\end{picture}}
\caption{Observed amplitudes of magnitude variations {\sl vs} logarithm of the "spin-down flux" $\dot{E} P_{orb}^{-4/3}$ for different BWs from Table~\ref{MSPs}. Squares correspond to the observed amplitudes, and  triangles  are their values corrected for the systems' inclinations. $P_{orb}$ is given  in hours and $\dot{E}$ in $10^{34}$ ergs s$^{-1}$.  The black rectangle corresponds to \psr. }
\label{fig4}
\end{figure*}

To estimate the dependence of the best solution  (Table~\ref{model}) on the fit parameters, we calculated the variation of the $\chi^2$ functional in the 2D planes (par1, par2) when the other parameters were fixed at the best values (see Figure~\ref{fig3}). The grey colour shows the values of  $y \equiv 1/\chi^2$  in the z-scale of 
$z = (y -y_\mathrm{min})/(y_\mathrm{max} - y_\mathrm{min})$. 
We define the  $1\sigma$ error of the fit as 0.68 from the maxima of the 2D plots. It is marked by the long-dashed lines in Fig.~\ref{fig3}. 

The fit shows that the object is close to filling its Roche lobe. The obtained distance to the source is 3.94$\pm0.16$ kpc, which is in accord 
with the most recent DM distance estimation based on the YMW16 Galactic electron density model (Sect. \ref{Intro}). 
The maximum difference between the "day-side" and "night-side" temperatures of the secondary is about 3000~K.
 The latter is similar to what is  observed in other BW systems. The interstellar extinction derived from the fit, $E(B-V)=0.085^{+0.030}_{-0.005}$, is close to the value proposed by  \citet{2011ApJ...737..103S} at the given distance (see Sect.~\ref{Intro}) and it is lower than that predicted by the extinction$-$distance maps by \citet{2018MNRAS.478..651G}.
  The masses of the pulsar and its companion are $ 1.35^{+0.5}_{-0.05} M_{\sun}$ and $0.034\pm{0.011}M_{\sun}$, respectively. The latter one is only $\approx$ 36 times  the Jupiter mass.  
The system inclination angle  $77\pm 13\degr $ indicates that the radio eclipse 
is most probably related to ionised material escaping from the companion. In Fig.~\ref{fig:MassFunc}, we show the 1$\sigma$ error box of the  three parameters  in the $M_{NS} - M_2$ plane. 
Its sizes are dominated by uncertainties of the optical fit, but not minor  uncertainties of the mass function and the projected semi-major axis of the binary orbit  mentioned above.

Our best fit model parameters provide the efficiency of the companion heating by the pulsar   $\eta\approx 0.2$. This value is in agreement with 
 the upper limit on the  efficiency of  re-radiation $\lesssim 0.5$ proposed by \citet{1969AcA....19..245R}.
 Discussion of  heating mechanisms and their efficiency in  binary MSPs can be found in, e.g., \citet{2013ApJ...769..108B, 2014ApJ...795..115L, 2016ApJ...828....7R, 2016ApJ...823..105D} and  \citet{ 2017ApJ...845...42S}. 
  Efficiency   factors in these systems typically lie in a range of $\approx 0.1-0.3$.
This implies that the relativistic pulsar wind powered by the spin-down luminosity represents an effective factor of heating. 
However, some  objects, e.g., PSR J1810$+$1744, show $\eta > 1$, implying strong anisotropy and/or inhomogeneity of the  wind. 

In Table~\ref{MSPs} we accumulated the observed, derived and modelling parameters of all BW MSPs that were studied via time-resolved optical observations. A correlation  is generally expected  between the brightness variation of the secondary and a "spin-down flux"  defined as $\dot{E} P_{orb}^{-4/3}$,  
\begin{equation}
\Delta m  \propto \log \left(\frac{F_{d}}{ F_n}\right)   \propto \log \left(\frac{T^4_{d}}{T^{4}_n}\right)  \propto \log \left(\eta\frac{\dot{E}}{4\upi a^2}\right)  \propto \log (\dot{E} P_{orb}^{-4/3} ).  
\label{eq6}
\end{equation} 
In Fig.~\ref{fig4} we show the full amplitude of the optical variations {\sl vs} logarithm of the "spin-down flux" for BWs with
available time-resolved optical photometry (squares). The filled triangles show the total amplitude of the variation corrected for the proposed system inclinations from the cited publications. As  seen, there is no evident correlation between the "spin-down flux" and  $\Delta m \propto (T_\mathrm{d} / T_\mathrm{n})^4$  at a scale of about two orders of the "spin-down flux".  Therefore, the origin of the "day-side" heating appears to be more complicated than suggested by Eq~\ref{eq6}. 
 As the pulsar wind is anisotropic, it is very likely that the heating effect highly depends on the unknown pulsar spin axis inclination. Moreover, it is possible that it is not only the pulsar wind that 
drives the heating. 
Specific structures 
of the surface magnetic field and possible 
convective zones 
of the secondary can also affect 
the observed optical variations of different BW companions.

\section{Conclusion} 

 We have presented the optical identification and time-resolved multi-band photometry of the BW PSR J2052+1219.
The detected object shows a strong optical variability with the period $P=2.752 h$, which coincides with the system orbital period 
reported by \citet{2016ApJ...819...34C} and derived from  the radio timing data.  
We reproduced the object light curve using  the model of the heating of the companion star by the pulsar. 
As a result,  we independently estimated the distance to the system of $\approx$ 4~kpc which is in accord with the   
 dispersion measure distance based on  the Galactic electron density model by \citet{2017ApJ...835...29Y}. 
 The companion mass is only $0.034\pm{0.011}M_{\sun}$ or about 36 times the Jupiter  mass M$_{J}$. Its radius is $\approx0.15R_{\sun}$  or only by a factor of 1.5 larger than the Jupiter radius $R_{J}$. It is close to filling its Roche lobe, and it has a gradient of the surface temperature 
of about 3000~K between the side facing the pulsar  ($\sim 6500$K) and the back side ($\sim 3200$K). For comparison,  brown dwarfs with similar masses typically have radii   
of about R$_{J}$ \citep[see, e.g.,][]{2017AJ....153...15B} and lower effective  temperatures of $\lesssim$ 2800 K \citep{2014A&ARv..22...80H}. We can speculate that as the system evolves, the companion can transform  
into a brown dwarf or a planet.      
At the current stage, its larger size and temperature are unconditionally 
determined by the pulsar wind.   The presence of eclipse in the radio  together 
with the $R$-band upper limit at the minimum brightness orbital phase imply a high inclination of the system. This is supported   
by our estimation of the inclination angle of $77\degr\pm13$ following from the optical light curve fit. The estimated heating efficiency 
of the companion by the pulsar is $\eta$ $\approx$ 0.2,  similar to that observed in other BW systems.  Maximum deviations of
individual observational optical points from the model   
light curve reach 
$\approx$ 20 per cent.  They can be likely attributed to the intrinsic short-time variations of the companion caused by 
complex plasma behaviors  near the companion surface due to  anisotropy and/or inhomogeneity of the pulsar wind.
Overall, PSR J2052+1219 is of a particular interest for further  spectroscopic and fast photometric studies 
using 
large telescopes. 

 We note that when this paper was submitted, 
\citet{2019arXiv190800992D} have published optical light curves of the considered 
system independently obtained with Keck, SOAR, and MDM telescopes. Their fit results including the companion night side temperature and mass, and the distance to the system are  
consistent with the values obtained by us, while the inclination angle of the orbit is significantly lower, about 54\degr. 
Considering such a small inclination, it may be difficult to explain the pulsar eclipse accounting for a small radius of the companion.
\citet{2019arXiv190800992D} do not constrain from their fit $E(B-V)$, the day-side temperature, the radius of the companion, its Roche-lobe filling factor, and the mass of the NS, fixing it at $1.5M_{\sun}$.

\section*{Acknowledgements}
We are grateful to the anonymous referee for useful comments which improved the paper.  
We thank F.~Camilo,  H.~T.~Cromartie and M.~Roberts for providing the information from the radio timing measurements. J.S.D. was supported by the NASA Fermi program. SZ and AK acknowledge PAPIIT grant IN$-$100617 for resources provided towards this research. 
The work of DZ and AK was funded by RFBR according to the research project 18-32-20170. The work of YS  was partially supported  by the RFBR grant 16-29-13009.
Based upon observations carried out at the Observatorio Astron\'{o}mico Nacional on the Sierra San Pedro M\'{a}rtir (OAN-SPM), Baja California, M\'exico. 
This work has made use of data from the European Space Agency (ESA) mission
{\it Gaia} (\url{https://www.cosmos.esa.int/gaia}), processed by the {\it Gaia}
Data Processing and Analysis Consortium (DPAC,
\url{https://www.cosmos.esa.int/web/gaia/dpac/consortium}). Funding for the DPAC
has been provided by national institutions, in particular the institutions
participating in the {\it Gaia} Multilateral Agreement.
The Pan-STARRS1 Surveys (PS1) and the PS1 
public science archive have been made possible through contributions by the Institute for Astronomy, the University of Hawaii, 
the Pan-STARRS Project Office, the Max-Planck Society and its participating institutes, the Max Planck Institute for Astronomy, Heidelberg 
and the Max Planck Institute for Extraterrestrial Physics, Garching, The Johns Hopkins University, Durham University, the University of Edinburgh, 
the Queen's University Belfast, the Harvard-Smithsonian Center for Astrophysics, the Las Cumbres Observatory Global Telescope Network Incorporated, 
the National Central University of Taiwan, the Space Telescope Science Institute, the National Aeronautics and Space Administration under 
Grant No. NNX08AR22G issued through the Planetary Science Division of the NASA Science Mission Directorate, the National Science Foundation Grant No. AST-1238877, 
the University of Maryland, Eotvos Lorand University (ELTE), the Los Alamos National Laboratory, and the Gordon and Betty Moore Foundation.

\end{document}